\documentclass[useAMS,usenatbib,usegraphicx]{mn2e}
\usepackage{aas_macros,url,pdfpages}

\title[The Unexpected 2012 Draconid Meteor Storm]{The Unexpected 2012 Draconid Meteor Storm}
\author[Q.-Z.~Ye et al.]
  {Quanzhi~Ye$^1$\thanks{E-mail: qye22@uwo.ca},
  Paul~A.~Wiegert$^1$, Peter~G.~Brown$^{1,2}$, \newauthor Margaret~D.~Campbell-Brown$^1$, and Robert~J.~Weryk$^1$\\
  $^{1}$Department of Physics and Astronomy, The University of Western Ontario, London, Ontario, N6A 3K7 Canada\\
  $^{2}$Centre for Planetary Science \& Exploration, The University of Western Ontario, London, Ontario, N6A 5B8 Canada}
\begin{document}

\date{Accepted 1970 January 1. Received 1970 January 1; in original form 1970 January 1}

\pagerange{\pageref{firstpage}--\pageref{lastpage}} \pubyear{2013}

\maketitle

\label{firstpage}

\begin{abstract}
An unexpected intense outburst of the Draconid meteor shower was detected by the Canadian Meteor Orbit Radar (CMOR) on October 8, 2012. The peak flux occurred at $\sim$16:40 UT on October 8 with a maximum of $2.4\pm0.3~\mathrm{hr^{-1}\cdot km^{-2}}$ (appropriate to meteoroid mass larger than $10^{-7}~\mathrm{kg}$), equivalent to a ZHR$_{\mathrm{max}}\approx9000\pm1000$ using 5-minute intervals, using a mass distribution index of $s=1.88\pm0.01$ as determined from the amplitude distribution of underdense Draconid echoes. This makes the outburst among the strongest Draconid returns since 1946 and the highest flux shower since the 1966 Leonid meteor storm, assuming a constant power-law distribution holds from radar to visual meteoroid sizes. The weighted mean geocentric radiant in the time interval of 15--19h UT, Oct 8, 2012 was $\alpha_g=262.4^{\circ}\pm0.1^{\circ}$, $\delta_g=55.7^{\circ}\pm0.1^{\circ}$ (epoch J2000.0). Visual observers also reported increased activity around the peak time, but with a much 
lower 
rate (ZHR$\sim200$), suggesting that the magnitude-cumulative number relationship is not a simple power-law. Ablation modeling of the observed meteors as a population does not yield a unique solution for the grain size and distribution of Draconid meteoroids, but is consistent with a typical Draconid meteoroid of $m_{total}$ between $10^{-6}$ to $10^{-4}$ kg being composed of 10 -- 100 grains. Dynamical simulations indicate that the outburst was caused by dust particles released during the 1966 perihelion passage of the parent comet, 21P/Giacobini-Zinner, although there are discrepancies between the modelled and observed timing of the encounter, presumably caused by approaches of the comet to Jupiter during 1966--1972. Based on the results of our dynamical simulation, we predict possible increased activity of the Draconid meteor shower in 2018, 2019, 2021 and 2025.
\end{abstract}


\section{Introduction}

The October Draconid meteor shower (DRA or IAU/MDC 009; usually referred to as ``Draconids''; some older literature may also refer to it as the ``Giacobinids'') is an annual meteor shower produced by comet 21P/Giacobini-Zinner and active in early October. It was first observed in the 1920s \citep{dav15,den27} and produced two spectacular meteor storms in 1933 and 1946 \citep[c.f.][]{jen95}. Aside from the two storms and a few outbursts, it has been quiet in most years, with hourly rates no more than a few meteors per hour.

A review of the historic observations and studies of the Draconids may be found in our earlier work \citep{ye13}; in this introductory section we mainly address the strong returns in 2011--2012. The outburst of the 2011 Draconids was predicted by a number of researchers \citep[][to name a few]{mas11,vau11}, and analyses made by various observing teams indicate that the predictions were quite accurate in both timing and meteor rates \citep[such as][and many others]{ker12,kot12,sat12,vau12}.

In contrast, no significant Draconid outburst was predicted for 2012. \citet{mas11} suggested encounters with the material ejected by Comet Giacobini-Zinner in 1959 and 1966 (in the form of the 1959- and 1966-trails), at 15--17h UT on Oct. 8, 2012, but he noted that ``visually-detectable activity is unlikely'', predicting the maximum Zenith Hourly Rate (ZHR) to be 0.5 and 0.2 respectively for each encounter. The prediction by Maslov is summarized in Table~\ref{tbl-maslov}.

In fact, an intense outburst of the Draconid meteor shower was subsequently detected by the Canadian Meteor Orbit Radar (CMOR) on Oct. 8, 2012 \citep{bro12}. As seen by CMOR, the meteor rate started increasing at $\sim 15$h UT, quickly reached a maximum around 16:40 UT, and returned to the background rate around 19h UT. Preliminary analysis indicated that the peak ZHR was well above the storm threshold (1 000 meteors per hour). Unfortunately, this was mostly a radar event; in addition to the poor timing of the peak (which only favored Siberia and Central Asia in terms of darkness), visual and video observers were largely caught unprepared. The quick look analysis carried out by the International Meteor Organization (IMO) revealed a peak of ZHR=$324\pm66$ centered at 16:51 UT, based on observations from only 4 observers\footnote{\url{http://www.imo.net/live/draconids2012/}.}. The camera set up by the Petnica Meteor Group in Serbia also detected some activity at 17--18h UT\footnote{\url{http://www.
meteori.rs/wordpress/i-ove-godine-pojacana-aktivnost-drakonida/}.}.

\begin{table*}
 \centering
  \caption{The Draconid 2012 outburst prediction by \citet{mas11}.}
  \begin{tabular}{ccccc}
  \hline
   Trail & Time & $\alpha_g$, $\delta_g$ & $v_g$ & ZHR \\
         & (Oct. 8, UT) & (J2000) & ($\mathrm{km \cdot s^{-1}}$) & \\
 \hline
1959	& 16:22, 16:54 & $262.7^{\circ}$, $+55.8^{\circ}$ & 21.0 & 0.5 \\
1966	& 15:37 & - & - & 0.2 \\
\hline
\end{tabular}
\label{tbl-maslov}
\end{table*}

Due to the lack of observations by other techniques, the radar observations are essential for the study of this outburst. We will mostly follow the methodology used in our earlier study of the 2011 event \citep{ye13} to analyze the 2012 event as recorded by CMOR. The goals of the present paper are to estimate the basic observational characteristics of the outburst, such as mean radiant, mass distribution, flux variation, and to apply the \citet{cam04} meteoroid ablation model to try to constrain the physical characteristics of the meteoroids. We will also investigate the cause of this outburst through numerical modeling of the stream.

\section{Instrumentation and Data Reduction}

The instrumental and data reduction details of this work is similar to that given in \citet{ye13} as applied to CMOR observations of the 2011 Draconids. Technical details of the entire system may also be found in \citet{hoc01,jon05,bro08} and \citet{wer12}. Here we only summarize a few key concepts.

CMOR is a six-site interferometric radar array located near London, Ontario, Canada. It operates at 17.45, 29.85 and 38.15 MHz, but only the 29.85 MHz data is used in this study. The 29.85 MHz component transmits with 12 kW peak power with a pulse repetition frequency of 532 Hz. The radar ``sees'' a meteor by detecting the reflection of the outgoing beam from an ionized meteor trail; therefore radar meteors are detected $90^{\circ}$ away from their radiants (such a reflection condition is called $specular$). As such, when the radiant is close to the zenith, it becomes difficult for the radar to detect echoes from a particular stream as the echoes are at low elevation and large ranges when the radiant is close to the zenith. Detected echoes are processed in an automatic manner to eliminate questionable detections, correlate the same events observed at different sites, and calculate meteor trajectories. Usually, raw streamed data is not permanently kept, but we intentionally saved the raw data from the 2012 
Draconid outburst for further analysis.

\begin{figure}
\includegraphics[width=0.5\textwidth]{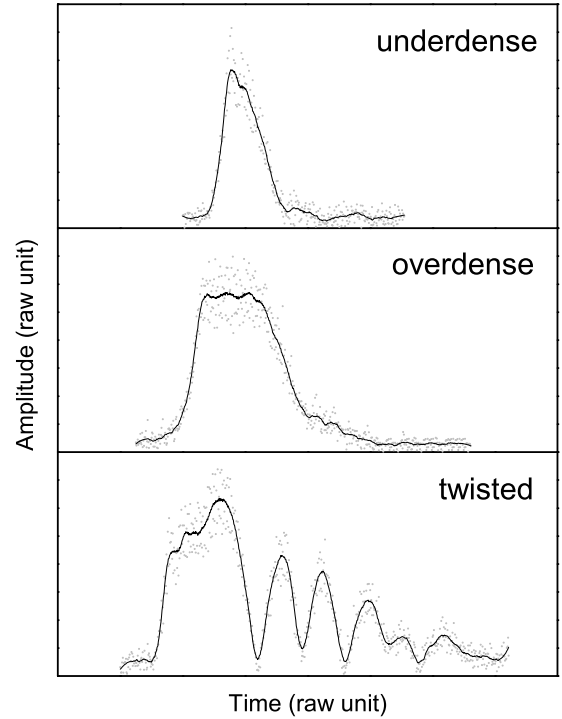}
\caption{Examples of meteor echoes as seen by CMOR. Top: underdense trail; middle: overdense trail; bottom: overdense, wind-twisted trail.}
\label{fig-example}
\end{figure}

Meteor echoes seen by radar may be classified as either $underdense$ or $overdense$ echoes, based on their apparent amplitude-time characteristics. The deciding factors are many \citep[see][for a more theoretical introduction]{ye13}, but for a given shower, overdense echoes are typically associated with larger meteoroids than underdense echoes. The underdense echoes are always characterized by a rapid rise and an exponential decay in amplitude (Figure~\ref{fig-example}a), making them relatively easy to identify automatically. Ideally, the overdense echoes are characterized by a rapid rise, a plateau, and a steady decay in amplitude (Figure~\ref{fig-example}b), but in reality trails distorted by upper atmospheric winds and exhibiting multiple reflection points are often seen (Figure~\ref{fig-example}c), making them difficult to characterize with automatic algorithms.

\begin{figure}
\includegraphics[width=0.5\textwidth]{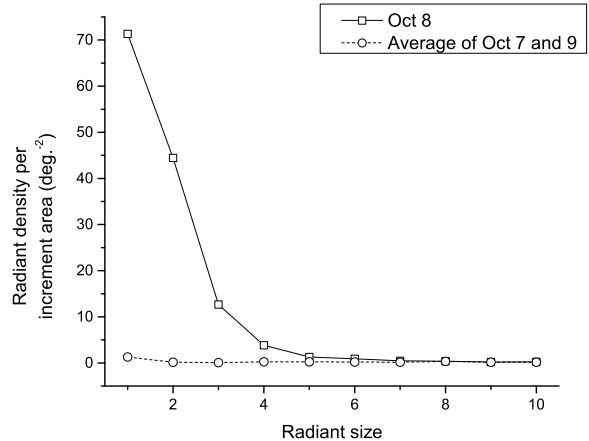}
\caption{Radiant densities centered at $\lambda-\lambda_{\odot}=53^{\circ}$, $\beta=+79^{\circ}$ in sun-centered ecliptic coordinates, as a function of radiant size on the outburst date, Oct. 8, 2012. Shown for comparison is the average of the same ecliptic sun-centered radiant location on Oct. 7 and Oct. 9, 2012 when no significant Draconid activity was present. It can be seen that the radiant size on Oct. 8 and the background rate meets at $\sim 5^{\circ}$.}
\label{fig-probe}
\end{figure}

Noting the advantages and weakness of the automatic routines, we take two approaches for the initial data reduction:

\begin{enumerate}
  \item For the first approach, we try to separate Draconid echoes from other echoes in the automatically processed data. This is done in several steps. First, we select all meteor radiants measured with time-of-flight (\texttt{tof}) velocities within $10^{\circ}$ of the apparent common radiant (i.e. an acceptance radius of $10^{\circ}$ around the apparent center of the cluster of meteor radiants) and $20\%$ of the annual geocentric velocity, $v_g=20.4~\mathrm{km \cdot s^{-1}}$ \citep{jen06}. The velocity restriction is intentionally broad, reflecting the spread due to measurement error and the high decelerations of Draconid meteoroids, a consequence of their fragility \citep[e.g.][]{jac50,cam06}. All these echoes are manually inspected to remove data with problematic time picks (i.e. an improperly automatically chosen time of occurrence of the specular point) determined by the automatic algorithms and to ensure the trajectory solution is correct. We do not revise the time picks, as we need to keep the 
procedure repeatable by automatic algorithms in order to estimate the uncertainty. The 
uncertainties are estimated with a Monto Carlo synthetic echo simulator developed by \citet{wer12}, allowing the derivation of a weighted mean common radiant. We then explore the ideal radiant size for this outburst event by varying the acceptance radiant interval to search for the acceptance radius at which the number of radiants reaches an asymptote approaching the background level, which in this case is $\sim 5^{\circ}$ (Figure~\ref{fig-probe}); therefore, we reduce the acceptance radius from $10^{\circ}$ to $5^{\circ}$. We identify 576 Draconid echoes in this manner and refer to this dataset as the \verb"complete" dataset (Table~\ref{tbl-rad}).
  \item For the second approach, we try to separate all overdense Draconid echoes. As noted before, overdense echoes are sometimes difficult to identify automatically, therefore we separate these echoes by manually inspecting the raw data from the main site between 15--19h UT on Oct. 8, 2012. Trajectories of these meteor echoes are manually determined when remote site observations are available to determine whether they are Draconids. In all, some 240 overdense Draconids are identified this way and will be referred as the \verb"overdense" dataset.
\end{enumerate}

\begin{figure}
\includegraphics[width=0.5\textwidth]{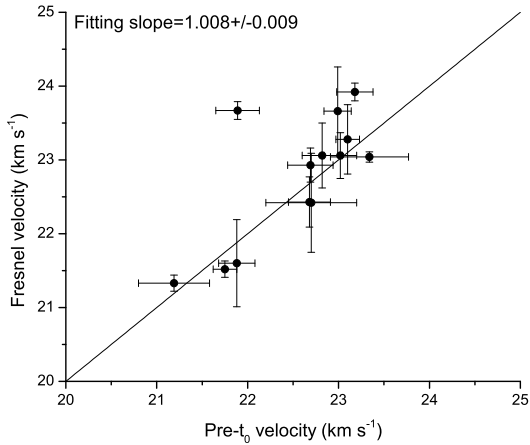}
\caption{The Fresnel velocities versus pre-t$_0$ velocities for the 13 high altitude echoes having both techniques for velocities measureable.}
\label{fig-ptn-fres}
\end{figure}

Since the velocities in the \texttt{complete} dataset are determined using the \texttt{tof} method, which depends on trail geometry and is uncertain in part due to height averaging, we also measure the speed of the meteoroids using the Fresnel phase-time method and Fresnel amplitude-time method \citep[simplified as ``pre-t$_0$'' and ``Fresnel'' method hereafter, see][\S4.6.1 and \S4.6.2 for an overview]{cep98}. The advantage of these two methods is that they only depend on the observation from the main site, which does not introduce the geometry or height averaging effects of the \texttt{tof} technique. In total, we gathered 360 echoes with pre-t$_0$ speeds and 25 echoes with Fresnel speeds, where 13 echoes have both pre-t$_0$ and Fresnel speeds. Numerically, the two speeds are equivalent, as the deceleration of the meteoroid is minimal during its passage over the Fresnel zones (amounting to $\sim1-2$ km of trail length; Figure~\ref{fig-ptn-fres}); we therefore assume the pre-t$_0$ or Fresnel speed is 
the meteoroid's speed at its 
specular height. For the 13 echoes where both pre-t$_0$ and Fresnel velocities are measured, preference is given to the one with smaller uncertainty.

\begin{figure}
\includegraphics[width=0.5\textwidth]{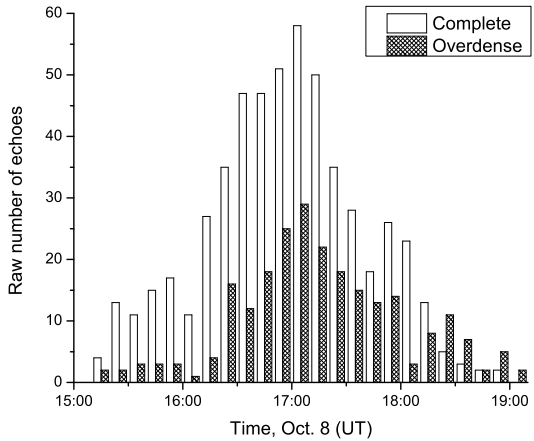}
\caption{The raw number of Draconid echoes detected by CMOR in the \texttt{complete} and the \texttt{overdense} dataset, binned in 10-minute intervals. The background level is normally $\ll1$ Draconid per 10-minute bin.}
\label{fig-num}
\end{figure}

The raw number of echoes as a function of time on Oct 8, 2012 in each dataset is given in Figure~\ref{fig-num}.

\section{Observational Results}

\subsection{Radiant and Orbit}

\begin{figure}
\includegraphics[width=0.5\textwidth]{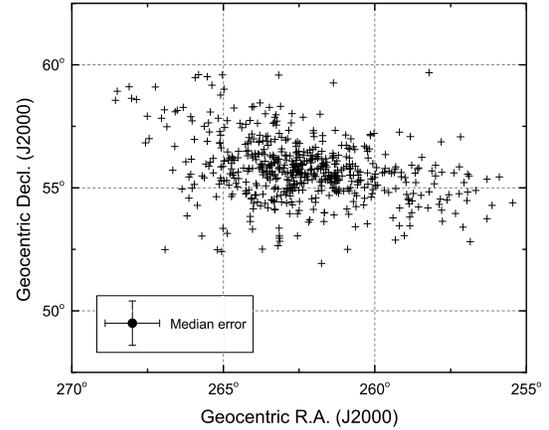}
\caption{Distribution of individual Draconid radiant of the \texttt{complete} dataset.}
\label{fig-rad}
\end{figure}

The weighted mean radiant of all meteors in the the \verb"complete" dataset is $\alpha_g=262.3^{\circ} \pm 0.1^{\circ}$, $\delta_g=+55.7^{\circ} \pm 0.1^{\circ}$ (J2000 epoch) (Figure~\ref{fig-rad}). The weighted mean radiant of meteors detected within 15--19h UT is $\alpha_g=262.4^{\circ} \pm 0.1^{\circ}$, $\delta_g=+55.7^{\circ} \pm 0.1^{\circ}$ (J2000 epoch).

% According to Maslov's prediction, two trails from different ejection epochs were predicted to contribute to the activity in 2012, so we also create two subsets for 15--16h UT and 16--17.5h UT to see a possible radiant shift due to different cometary trails, which revealed a poleward shift of $0.4^{\circ} \pm 0.2^{\circ}$ between these two analysis periods (Table~\ref{tbl-rad}).

% The geocentric mean velocity determined from the \verb"complete" dataset is $18.50 \pm 0.06~\mathrm{km \cdot s^{-1}}$, this is $\sim 10\%$ lower than the value associated with annual (non-outburst) returns of the shower ($20.4~\mathrm{km \cdot s^{-1}}$). In the 15--16h UT and 16--17.5h UT subsets, a decline of $v_g$ can be seen. The likely cause, as noted by \citet{cam06} and \citet{ye13}, is instrumental rather than physical: the fragility of Draconids causes them to decelerate more significantly than normal meteors, which resulting an underestimation of the deceleration correction. The magnitude of underestimation also depends on radiant elevation, as high radiant elevation will leads to a deeper penetration through the atmosphere (Figure~\ref{fig-ht-time}) and results a more significant deceleration. As the Draconid radiant was rising during the outburst hours, it makes sense that the deceleration is more significant in later hours.

\begin{table*}
 \centering
  \caption{Characteristics of the 2012 Draconid radiant as observed by CMOR. The mean radiants are calculated using error weightings according to the Monte Carlo error simulations by the synthetic echo simulator \citep[see \S 2.1 of this paper, as well as][]{wer12,ye13}.}
  \begin{tabular}{lccc}
  \hline
   Time period & $\alpha_g$, $\delta_g$ & $N_{radiants}$ \\
    & (J2000) & \\
 \hline
 Oct. 8 (all day) & $262.32^{\circ}\pm0.10^{\circ}$, $+55.67^{\circ}\pm0.05^{\circ}$ & 576 \\
 Oct. 8 (15--19h UT) & $262.35^{\circ}\pm0.10^{\circ}$, $+55.67^{\circ}\pm0.05^{\circ}$ & 545 \\
% \hline
% Oct. 8 15:00--16:00 UT & $262.15^{\circ}\pm0.28^{\circ}$, $+55.37^{\circ}\pm0.12^{\circ}$ & 68 \\
% Oct. 8 16:00--17:30 UT & $262.44^{\circ}\pm0.13^{\circ}$, $+55.75^{\circ}\pm0.06^{\circ}$ & 361 \\
\hline
\end{tabular}
\label{tbl-rad}
\end{table*}

% \begin{figure}
% \includegraphics[width=\textwidth]{ht-time.eps}
% \caption{The change of echo specular height with respect to time. It is apparent that higher radiant elevation will leads to deeper penetration through the atmosphere, hence the deceleration is more significant for echoes detected in the later hours.}
% \label{fig-ht-time}
% \end{figure}

The derivation of true orbit requires additional effort in data reduction, as the fragility of Draconid meteoroids introduces an underestimation of the deceleration correction as noted by \citet{cam06} and \citet{ye13}, that leads to a reduction in estimating the true out of atmosphere speed of the meteoroids. To solve this problem, we make use of the echoes detected at higher heights, as they do not penetrate very deep into the atmosphere and therefore suffer less deceleration. We also need to select the echoes with pre-t$_0$ or Fresnel velocity measured to minimize the error caused by deceleration undercorrection. We only consider the echoes detected within 15--19h UT (i.e. close to the peak time) to minimize the possible contamination of background Draconid meteors not associated with the outburst. We then choose 98 km as the cut-off height, as it is the highest possible cut-off with a statistically significant sample (only two echoes are detected at $\geq99$ km). Taking all these constraints into account,
 a total of 14 
echoes are 
selected from the \texttt{complete} dataset. We note that a significant fraction ($5/14$) of these have local zenith angle $\eta$ above $50^{\circ}$, which suggests that the undercorrection effect is again significant due to the longer atmospheric path of the meteoroid entering the atmosphere at such a shallow angle. Hence, we remove these five echoes and create an additional subset, and present both results together (Table~\ref{obs-orbit}).

\begin{table*}
\vbox to220mm{\vfil Landscape table to go here.}
\caption{}
\label{obs-orbit}
\end{table*}

We note that the mean orbit derived from the nine-meteor sample is consistent with the orbit of 21P/Giacobini-Zinner at its 1959 and 1966 apparitions. Additionally, the $v_g$ derived from the nine-meteor sample agrees with that derived from the multi-station video observations of the 2011 Draconid outburst \citep[$v_g=20.9\pm1.0~\mathrm{km \cdot s^{-1}}$ as noted by][]{jen11} within uncertainty. This agreement is reassuring as the video observations have more complete coverage of the trail length of each meteoroid compared to radar, allowing the deceleration correction for individual meteoroids to be established with higher confidence along with the pre-atmosphere velocity ($v_g$) and orbit. We believe that the orbit derived from the nine-meteor sample represents the best mean orbit for the 2012 Draconid outburst as observed by CMOR.

\subsection{Mass Index}

The mass index can be determined using either underdense echoes or overdense echoes. For underdense echoes, the cumulative number of echoes with amplitude greater than $A$ follows the relation $N \propto A^{-s+1}$ \citep{mci68}, while for diffusion-limited overdense echoes, $N \propto \tau^{-\frac{3}{4}(s-1)}$ \citep{mci68}, where $\tau$ is the echo duration.

\begin{figure}
\includegraphics[width=0.5\textwidth]{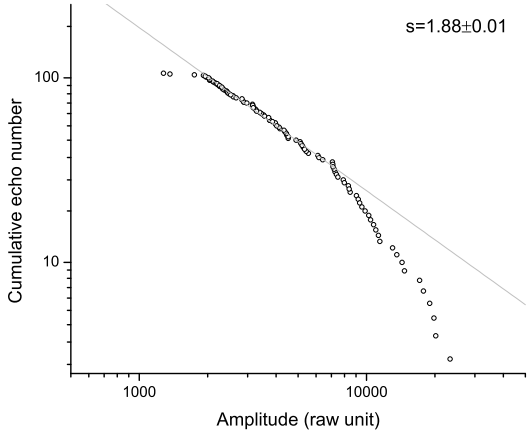}
\caption{Cumulative echo numbers as a function of amplitude for Draconid echoes with ranges between 110--130km. The mass index is determined to be $1.88 \pm 0.01$ by fitting the linear portion of the curve as shown.}
\label{fig-ud-amp}
\end{figure}

We first use the underdense echoes detected between 15--19h UT in the \texttt{complete} dataset to determine $s$. Echoes within a range interval between 110--130 km from the main site are selected to avoid contamination from overdense-transition echoes \citep[see][for discussion]{bla11}. We determine the mass index to be $1.88 \pm 0.01$ by fitting the linear portion of the data in Figure~\ref{fig-ud-amp}. The uncertainty given here is the fitting uncertainty only, and could be several times smaller than the real one due to the small sample size.

\begin{figure}
\includegraphics[width=0.5\textwidth]{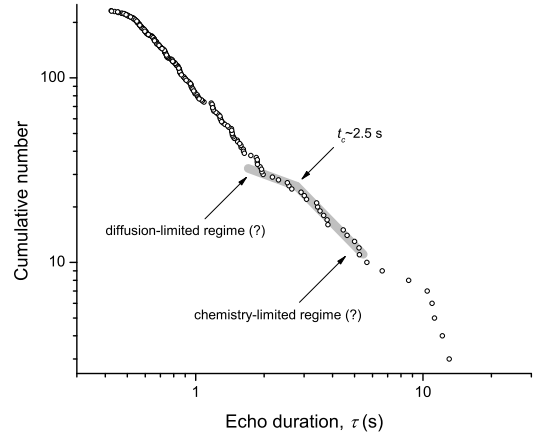}
\caption{Cumulative echo numbers as a function of echo duration for overdense Draconid echoes. The possible characteristic time can be seen near $t_c=2.5$ s.}
\label{fig-ov-tc}
\end{figure}

\begin{figure}
\includegraphics[width=0.5\textwidth]{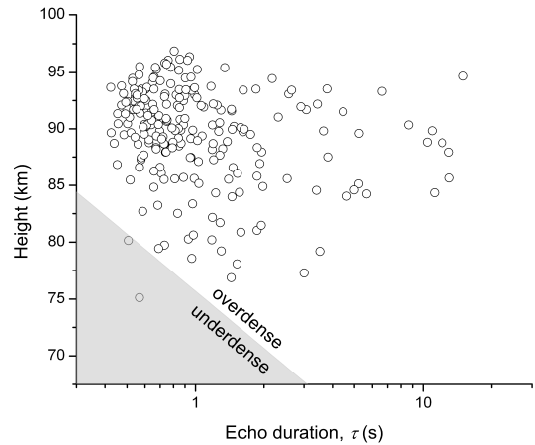}
\caption{The height range of the selected overdense echoes in the \texttt{overdense} dataset. The shaded area marks the underdense region defined by \citet{mck61}. If this population
were mainly underdense, a clear duration vs. height trend would be presented in the graph.}
\label{fig-ht-dur}
\end{figure}

We then use the \texttt{overdense} dataset as a check on the mass index value determined using the amplitude distribution alone (Figure~\ref{fig-ov-tc}). The trail of electrons can fall to an undetectable density either through diffusion or chemical recombination. The latter is more likely for very dense trails produced by larger meteoroids at low altitudes. The possible turnover between diffusion- and chemistry-limited regime, or the ``characteristic time'', can be seen at about 2.5 s in Figure~\ref{fig-ov-tc}, which agrees with the value derived for 2011 \citep[2.7s, ][]{ye13}. Using the few data points in the possible diffusion-limited regime, the mass index can be estimated to be around 1.7--1.8. However, we should note that this result is doubtful due to the presence of a sudden steep rise in cumulative number at $\tau<\sim2$ s. The possibility of underdense contamination can be ruled out by examining the theoretical underdense region in the height-duration distribution (Figure~\ref{fig-ht-dur}). A 
possible explanation of this behavior is the lack of long overdense echoes due to 
the radiant geometry at the time of the outburst as well as abundance of smaller meteoroids of this event, preventing us from getting enough statistics for longer duration echoes. 
Alternatively, the lack of a power law fit at the high mass end of the meteoroid distribution (i.e. overdense echoes) may indicate that the size distribution within the stream does not follow a power-law, in contrast to the behavior seen in 2011, where a clear fit existed to the smallest overdense echoes durations. In this case, the upper upturn below $\tau<\sim2$ indicates an overabundance of smaller Draconids in the outburst.

\subsection{Flux}

The flux can be calculated from the number of echoes detected per unit time divided by the effective collecting area of the radar for the Draconid radiant. The effective collecting area of CMOR is calculated following the scheme described in \citet{bro95}. Simply put, the collecting area for a given radiant is the integration of the magnitude of the gain over the ``reflecting strip'' of the radar wave (a $90^{\circ}$ great circle perpendicular to the radiant direction), which takes into account both the radiant geometry and mass index. The ZHR (i.e. the number of meteors that an average observer would see in one hour, given that the sky is clear and dark, and the radiant is at the zenith) can then be calculated by using its relationship to the actual meteoroid flux \citep{kos90} as computed by the radar and mass index, using the fact that the limiting meteor radio magnitude of CMOR is $\sim8.5$.

\begin{figure*}
\includegraphics[width=\textwidth]{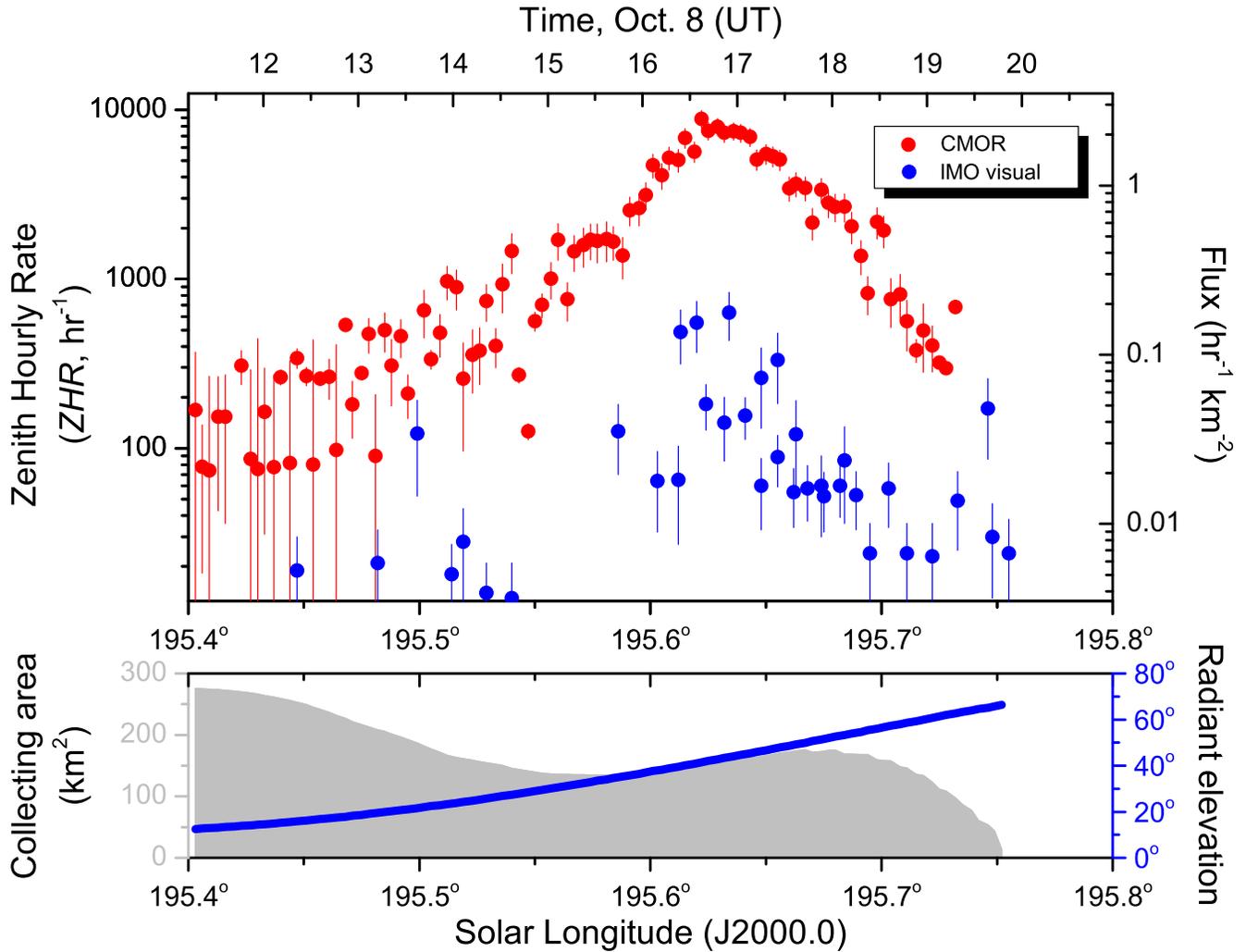}
\caption{The ZHR/flux profile of the 2012 Draconid outburst as observed by CMOR (binned in 5-minute intervals; error bars represent Poisson errors only), comparing to the visual profile reported by the visual observers to the International Meteor Organization (IMO). The radar flux is appropriate for meteoroids with $m>10^{-7}~\mathrm{kg}$; the visual flux is appropriate for meteors with $V<2-3~\mathrm{mag}$.}
\label{fig-flux}
\end{figure*}

As shown in Figure~\ref{fig-flux}, the main peak based on 5-min averaged CMOR data occurred at 16:38 UT (solar longitude, $\lambda_{\odot}=195.622^{\circ}$), Oct. 8, with maximum flux of $2.4\pm0.3~\mathrm{hr^{-1}\cdot km^{-2}}$ (appropriate to meteoroid mass larger than $10^{-7}~\mathrm{kg}$), equaling ZHR$_{\mathrm{max}}=9000\pm1000$. The IMO data shows several ``peak-lets'' from 16:25--16:55 UT, with ZHR$_{\mathrm{max}}$ around 500--600. The existence of these ``peak-lets'' is likely a statistical oddity due to the limited number of observers (only one observer for each of these ``peak-lets''). For example, the time range around 16:40 UT (which contained three observers reporting 10 Draconids total observed) corresponds to a ZHR of $185\pm55$.

We note that these ZHRs assumed a single power-law fit from fainter radar meteors to equivalent radio magnitude of +6.5 where the ZHR is defined. As noted earlier, an apparent deviation is notable between larger visual meteoroids and smaller radar meteoroids from a pure power-law, hence the effective ZHR quoted from CMOR observations are likely upper limits.

\subsection{Ablation Modeling}

To model the structure of the Draconid meteoroids, we used the meteoroid ablation model developed by \citet{cam04}. The velocities and electron line densities as measured by CMOR are used as constraints to fit the model. The velocities used here are appropriate to the echo specular height (i.e. pre-t$_0$ or Fresnel velocities); the electron line densities are computed from the observed amplitude value, using the formula appropriate to either underdense echoes (when $q<2.4\times10^{14}~\mathrm{m^{-1}}$) or overdense echoes (when $q>2.4\times10^{14}~\mathrm{m^{-1}}$). The methodology is essentially the same as for the 2011 outburst \citep{ye13}, except that we do not model individual events for deceleration as no such events are found for the 2012 outburst. We use the parameters suggested by two previous studies of the Draconids, \citet{borspukot07} and \citet{ye13}, while other parameters are left fixed as suggested in \citet{cam04}. The parameters are summarized in Table~\ref{metsim-par}.

\begin{table*}
 \centering
  \caption{Input parameter for the ablation model as used in \citet{ye13}.}
  \begin{tabular}{lcl}
  \hline
  Parameter & Value & Unit \\
  \hline
Deceleration corrected apparent velocity, $v_{\infty}$ & 23.27 & $\mathrm{km \cdot s^{-1}}$ \\
Zenith angle, $\eta$ & $55^{\circ}$ & deg. \\
Bulk density, $\rho_{bulk}$ & 300 & $\mathrm{kg \cdot m^{-3}}$ \\
Grain density, $\rho_{grain}$ & 3 000 & $\mathrm{kg \cdot m^{-3}}$ \\
Heat of ablation, $q_{heat}$ & $3 \times 10^6$ & $\mathrm{J \cdot kg^{-1}}$ \\
Thermal conductivity, $\kappa$ & 0.2 & $\mathrm{J \cdot m^{-1} \cdot s^{-1} \cdot K^{-1}}$ \\
\hline
\end{tabular}
\label{metsim-par}
\end{table*}

\begin{figure}
\includegraphics[width=0.5\textwidth]{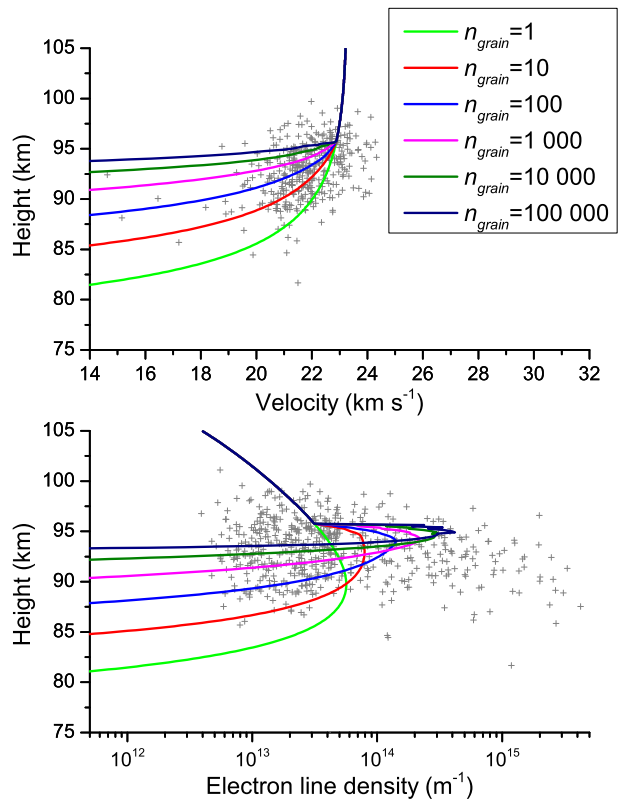}
\caption{Sensitivity test for different number of grains, $n_{grain}$, at a fixed total mass $m=10^{-5}~\mathrm{kg}$.}
\label{fig-sensitivity1}
\end{figure}

We first tried to find the optimal combination of grain mass and grain number by fixing the total mass at $10^{-5}$ kg and varying these two variables. We choose $10^{-5}$ kg here, as this value is a reasonable compromise between the detection limit ($\sim10^{-7}$ kg) and the underdense/overdense transition level ($\sim10^{-4}$ kg appropriate to meteors with $q\approx 10^{14}~\mathrm{m^{-1}}$ or peak visual brightness of +4 magnitude). We used the transition level as our upper limit as most of the echoes in the \texttt{complete} dataset are underdense echoes. Finally, we look for a modeling fit that minimizes the deviation between the model and trend of the data points with respect to the height. However, from Figure~\ref{fig-sensitivity1}, we cannot identify a unique fit; the range of optimal fit seems to lie somewhere between $n_{grain}=10$ and $n_{grain}=10~000$.

\begin{figure*}
\includegraphics[width=\textwidth]{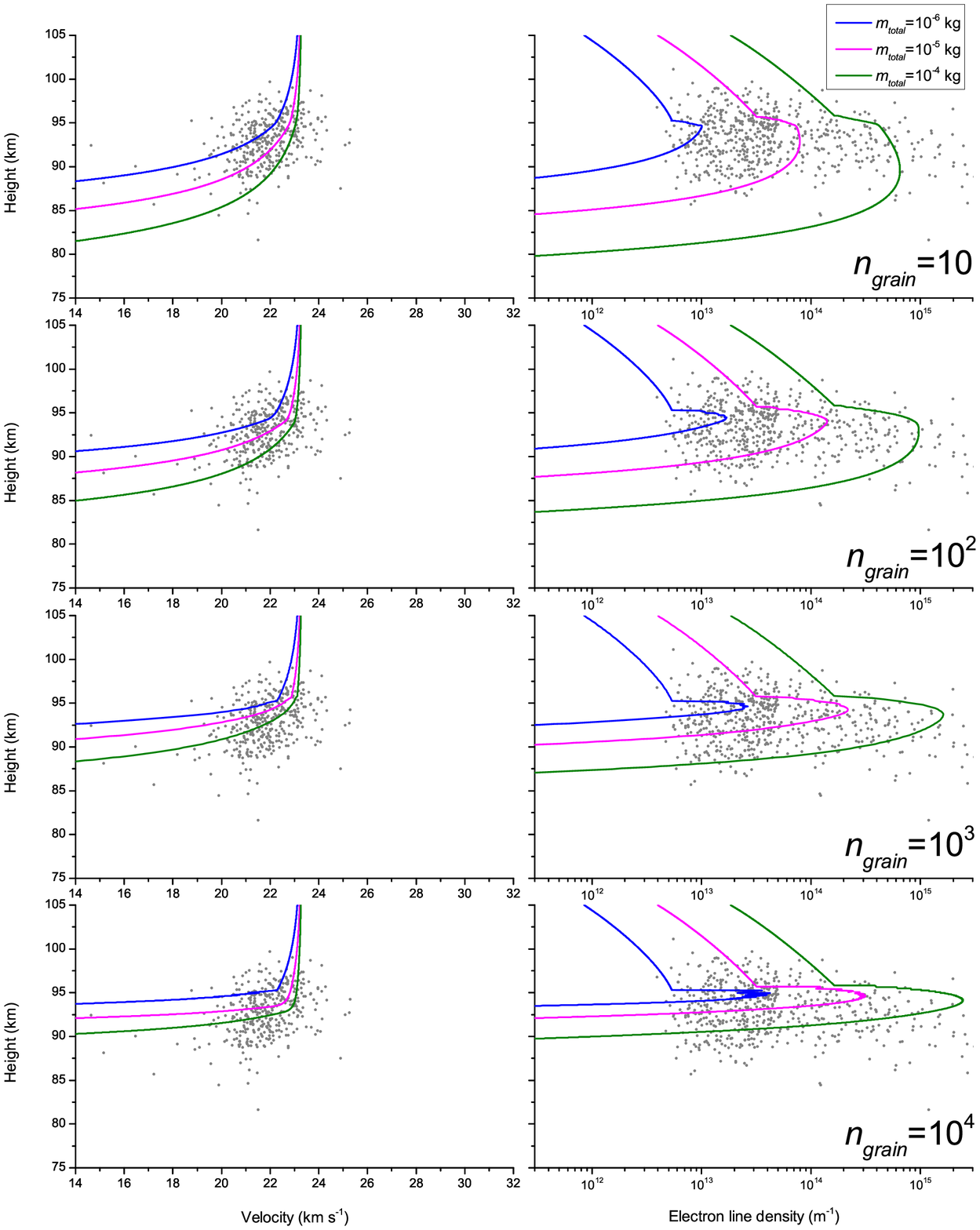}
\caption{Sensitivity test for different number of grains, $n_{grain}$ but for varying $m_{grain}$ (ranges from $10^{-10}$ to $10^{-5}$ kg). The corresponding $m_{total}$ range is from $10^{-6}$ to $10^{-4}$ kg.}
\label{fig-sensitivity2}
\end{figure*}

We then plot the modeling fit for $m_{total}=10^{-6}$ to $10^{-4}$ kg, for $n_{grain}=10$ to 10 000, to further examine the goodness of each parameter set (Figure~\ref{fig-sensitivity2}). Again, we do not see an obviously unique solution, but it seems that the $n_{grain}=10$ and $n_{grain}=100$ scenarios produce a better qualitative fit to both velocities and electron line densities. This generally agrees with our previous finding for the 2011 outburst which find the optimal $n_{grain}=100$; it also agrees with the photographic result for the 2005 outburst to some extent \citep{borspukot07}, but we note that the fits are not well constrained from our observations, therefore they can only be used to broadly establish the fact that the 2012 Draconid meteoroids were not radically different in physical makeup from those detected during the 2011 outburst.

\section{Dynamic Modeling}

To better understand the 2012 Draconids outburst, numerical simulations were performed of the parent, comet 21P/Giacobini-Zinner. The simulations included a Solar System of eight planets whose initial conditions were derived from the JPL DE405 ephemeris \citep[]{sta98}, with the Earth-Moon represented by a single particle at the Earth-Moon barycenter. The system of planets, parent and meteoroids were integrated with the RADAU method \citep[]{eve85} with a time step of seven days used in all cases.

The comet orbital elements used in these simulations are derived from \citet{marwil08} where orbital elements are provided for each appearance of 21P/Giacobini-Zinner from its discovery in 1900 through 2005. This extensive data set is important as the parent, being a Jupiter-family comet, is extensively perturbed, and has time-varying non-gravitational parameters.

An initial survey of the meteoroid complex used the catalog orbit for 21P/Giacobini-Zinner's 2005 perihelion passage as a starting point. Simulations of meteoroids released during each of 21P/Giacobini-Zinner's perihelion passages back to 1862 were examined for clues as to the origin of the material that produced the 2012 outburst. These simulations revealed that the 1966 apparition was the source of the 2012 event, and as a result the orbital elements given for the 1966 perihelion passage were used in subsequent simulations. These orbital elements are listed in Table~\ref{ta:elements}.

\begin{table*}
\begin{tabular}{lccccccc} 
\hline
 Name     & Perihelion & $q$ (AU) & $e$ & a (AU) & $\omega$ & $\Omega$ & $i$ \\ \hline
  21P      & 2005 Jul 2.7605  & 1.037914 & 0.705691  & 3.526613 &  172.5429$^{\circ}$  & 195.4301$^{\circ}$ &  31.8109$^{\circ}$ \\
  21P      &1966 Mar 28.2844  & 0.933501 & 0.729394  & 3.449669 &  172.9153$^{\circ}$  & 196.6674$^{\circ}$ &  30.9382$^{\circ}$ \\
  sim. &    & 0.9968 $\pm$ 0.0003 & 0.725 $\pm$ 0.003 & 3.62 $\pm$ 0.04 & 172.75$^{\circ}\pm0.09^{\circ}$ & 195.65$^{\circ}\pm0.04^{\circ}$ & 31.79$^{\circ}\pm0.02^{\circ}$\\
\hline
  obs. &  & $0.9952\pm0.0003$ & $0.710\pm0.032$ & $3.22\pm0.61$ & $172.26^{\circ}\pm0.35^{\circ}$ & $195.620^{\circ}\pm0.001^{\circ}$ & $31.45^{\circ}\pm0.69^{\circ}$ \\
\hline
\end{tabular}
\caption{A list of the orbital elements used for comet 21P/Giacobini-Zinner for dynamic modeling of the Draconid meteoroid stream formation. From \citet{marwil08}. CMOR observations are extracted from the nine-meteor solution in Table~\ref{obs-orbit}.}
\label{ta:elements}
\end{table*}

No non-gravitational forces due to outgassing \citep[e.g.][]{marsekyeo73} were applied to the comet in our final simulation results. Some test simulations that did include non-gravitational forces, using the values for 21P/Giacobini-Zinner from \citet{yeo86} as listed in \citet{marwil08}, were run but these did not produce any noticeable differences in the results.

The simulations were run with the two-stage refinement procedure described in \citet{wiebrower13}. Briefly, in the first stage, the comet orbit is integrated backwards to the desired starting point, in this case back approximately 150 years to 1862. The comet is then integrated forward again, releasing meteoroids at each perihelion passage as it does so. All meteoroids which pass sufficiently close to the Earth's orbit during the simulation are collected: this is our list of ``bulls-eyes''. Bulls-eyes are those that have a minimum orbital intersection distance (MOID) of no more than 0.02 AU, and that are at their MOID with the Earth no more than $\pm 7$ days from when our planet is there.

In the second stage, the integration of the comet forward in time is repeated. However in this case, instead of releasing particles randomly as determined by a cometary ejection model, particles are released only near initial conditions in which a bulls-eye was produced in the same time step in the first simulation. These second generation particles are released at the same position (that of the nucleus, taken to be a point particle) but are given a random change in each velocity component of up to $\pm 10$\% of that of the original bulls-eye. These second generation meteoroids inevitably contain far more particles that reach our planet. Of this sample, those which pass closest to the Earth in space and time will be considered to constitute the simulation outburst.

In our initial simulations, at each perihelion passage a number $M=10^4 - 10^5$ particles is released in each of the four size ranges with radii $r$ from $10^{-5}-10^{-4}$~m, $10^{-4}-10^{-3}$~m, $10^{-3}-10^{-2}$~m and $10^{-2}-10^{-1}$~m, the whole range spanning from 10 $\mathrm{\mu m}$ to 10 cm in diameter.  They are chosen so that the distribution of particle radii is flat when binned logarithmically by size. A power-law size distribution with mass index $s$ can be recovered after the simulation is complete by giving each particle a weight proportional to $r^{-3s+3}$ \citep[]{wievaucam09} if desired. However, we do not apply such a weight to the results reported here because the outburst is found to consist of such a narrow range of particles (radii almost exclusively from 100 $\mathrm{\mu m}$ to 1~mm) that the application of such a weighting is likely inappropriate.

Post-Newtonian general relativistic corrections and radiative (i.e. Poynting-Robertson) effects are also included. The ratio of solar radiation pressure to gravitational force $\beta$ is related to the particle radius $r$ (in $\mu$m) through $\beta = 1.9/r$ following \citep{weijac93}, though our expression assumes a particle mass density $\rho = 300~\mathrm{kg \cdot m^{-3}}$ for Draconid meteoroids as was reported by \citet{borspukot07}.

The comet is considered active (that is, simulated meteoroids are released) when at a heliocentric distance of 3~AU or less, during the first simulation stage the simulated parent releases particles at a uniform rate during this part of its orbit. While active, particles are released with velocities from the prescription of either \citet{crirod97} or the revised Whipple
model of \citet{jon95}. We use a nucleus radius for 21P/Giacobini-Zinner of 1000~m \citep{tan00} but in the
absence of other details we assume a Bond albedo for the nucleus of 0.05, a nucleus density of 300~$\mathrm{kg \cdot m^{-3}}$ and an active fraction of the comet's surface of 20\% where needed in the above ejection models. The \citet{bro95} model was found to reproduce the duration of the outburst as observed by CMOR slightly better and the results reported here use the Jones model.

A supplementary integration of the comet orbit backwards for a thousand years allowed a determination of the Lyapunov exponent using the algorithm of \citet{mikinn99}. The e-folding timescale is approximately 30 years, not unexpected for a Jupiter-family comet with a node near that giant planet. Our primary result, that the outburst originated from the 1966 perihelion passage of the parent comet, is thus two e-folding times into the past and thus not unduely affected by chaotic effects.  The short Lyapunov time is a result of the frequent close encounters that 21P suffers with Jupiter. The one of most relevance here is the only close encounter occuring in the 1966-2012 time frame, a close approach to within less than 1.6 Hill radii in 1969. This encounter strongly affects both the parent and the meteoroids released during the 1966 perihelion passage.

\section{Comparison of simulations with observations}

\begin{figure*}
\includegraphics[width=\textwidth]{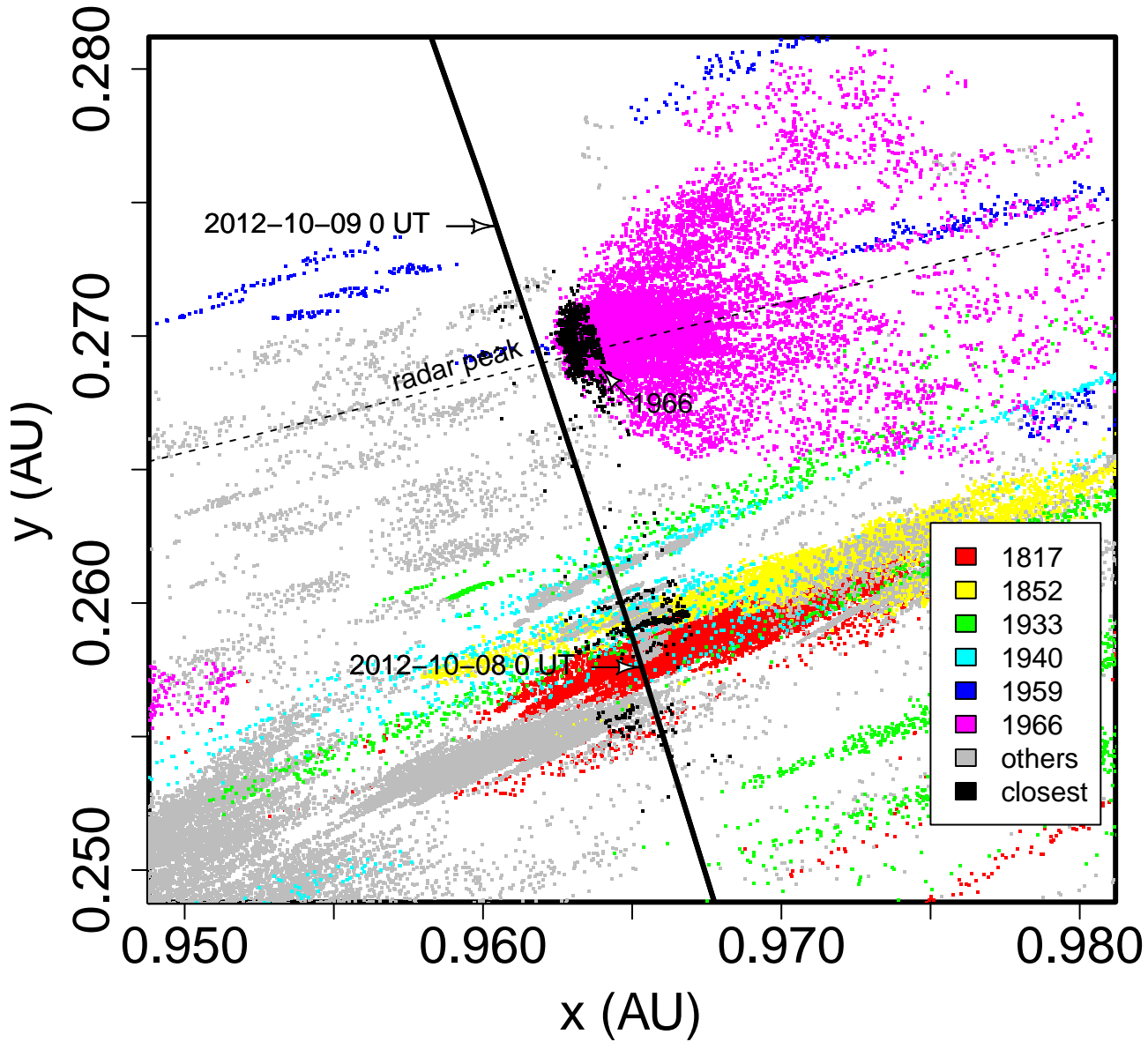}
\caption{The nodal footprint of the Draconid meteor stream for 2012 from simulation. The most populous perihelion passages are indicated by specific colors, all others are in grey. Black is reserved for those particles
which meet the most stringent encounter criteria and are deemed to comprise the simulated shower. Only the 1966 perihelion passage has a substantial number of such particles. The timing of the observed CMOR radar peak is
indicated by a dotted line. See text for more details.}
\label{fi:footprint}
\end{figure*}

The simulation output for the year 2012 is presented in Figure~\ref{fi:footprint}, which shows the nodal intersection points of the simulated meteoroids overplotted on the Earth's orbit. The dots shown are those test particles within 0.02~AU of Earth's orbit and reaching that point within $\pm7$~days of our planet being at the same solar longitude. These are colour-coded by the perihelion of the parent that released them. The black points represent those which are closest to intersecting the Earth, within $\Delta r = \pm 0.002~\mathrm{A.U.}$ and $\Delta t = \pm1$~day; these will be taken to constitute the simulated outburst.

Only one perihelion passage has a significant population of particles that meet the outburst criteria: 1966. The location of the peak corresponds to the time that CMOR detected the outburst, though the nodal footprint is slightly off the Earth's orbit: this will be discussed further below. 

\subsection{Why was there no prediction of an outburst?}

The largest difference between the simulations and observations is the fact that the simulated meteors approach but do not quite intersect the Earth's orbit. This likely contributed to the absence of theoretical predictions of the outburst. Maslov\footnote{\url{http://feraj.narod.ru/Radiants/Predictions/1901-2100eng/Draconids1901-2100predeng.html}.} did predict activity for the 2012 Draconids at this time, in particular in the radio meteor range, which he attributed to the 1959 and 1966 trails, which our simulations confirm.

In order to investigate the question of why the simulated nodal footprint of the 1966 streamlet does not intersect the Earth's orbit, we performed additional simulations with higher ejection velocities to determine the conditions required for these meteoroids to reach the Earth. Additional ejection velocities around 50--100~$\mathrm{m \cdot s^{-1}}$ are needed to move the simulated 1966 stream to Earth orbit crossing. Particles in our model are ejected with a range of velocities
$V_{ej}$ which depends on their size and heliocentric distance. Particles from $10^{-5}$ to $10^{-4}$ m have $V_{ej} = 210 \pm
80~$m/s (average $\pm$ one standard deviation); $10^{-4}$ to $10^{-3}$ m, $V_{ej} = 66 \pm 25$ m/s;  $10^{-3}$ to $10^{-2}$ m, $V_{ej} = 21 \pm 8 $ m/s and $10^{-2}$ to $10^{-1}$ m = $6.6 \pm 2.5$ m/s. The three largest size ranges can be brought to Earth intersection by increasing the ejection velocity to 50 to 100~m/s, while 30~m/s proves just insufficient. Thus the discrepancy in the nodal position (which amounts to roughly 0.001~A.U. or about half the distance to the Moon) could be accounted for by relatively high ejection velocities (as Maslov also noted). However we note that the close encounter with Jupiter suffered by 21P/Giacobini-Zinner and its daughter particles in 1969, as well as uncertainties in the comet orbit itself (i.e. its non-gravitational parameters changed markedly in the 1966--1972 time frame, \citet[]{yeo86}) make it difficult to pinpoint a single cause. We conjecture that inevitable errors in the Jupiter family comet's orbit coupled with its frequent 
approaches 
to Jupiter produce uncertainties in the location of the stream's node large enough to confound attempts to model the shower accurately.

Nonetheless, we can conclude with some confidence that the outburst observed by CMOR was produced by the 1966 trailet. This is because a comparison of the properties of these particles are consistent to a high degree with the Draconid outburst observed by CMOR. This includes the timing of the outburst, its radiant, velocity and the overall orbital elements, which we present below.

\subsection{Timing and radiants}

\begin{figure}
\includegraphics[width=0.5\textwidth]{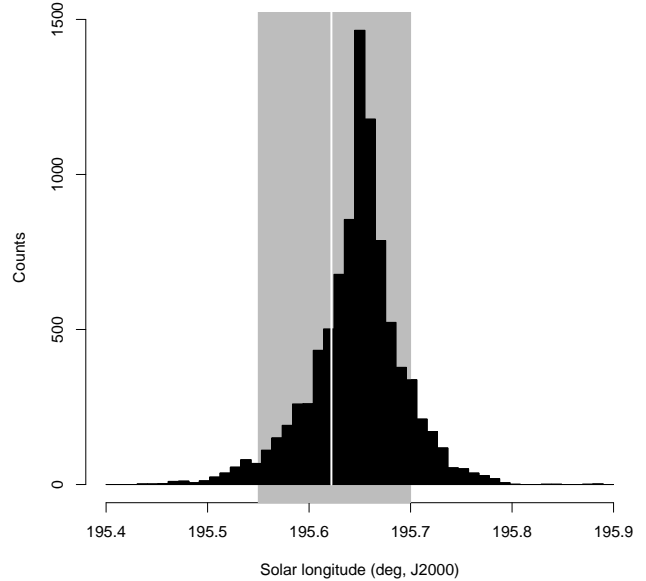}
\caption{The timing of the Draconid outburst of 2012. The black histogram shows the simulation; the grey band indicates the times where CMOR saw significant shower activity and the white line indicates the radar peak.}
\label{fi:sollon}
\end{figure}

Figure~\ref{fi:sollon} shows the timing of the shower. The radar and simulation results are offset slightly, with the simulations showing activity peaking about $0.03^{\circ}$ of solar longitude (about 45 minutes) earlier than was observed, though the precise width and peak of the simulated shower is somewhat sensitive to our choice of $\Delta r$ and $\Delta t$.

\begin{figure}
\includegraphics[width=0.5\textwidth]{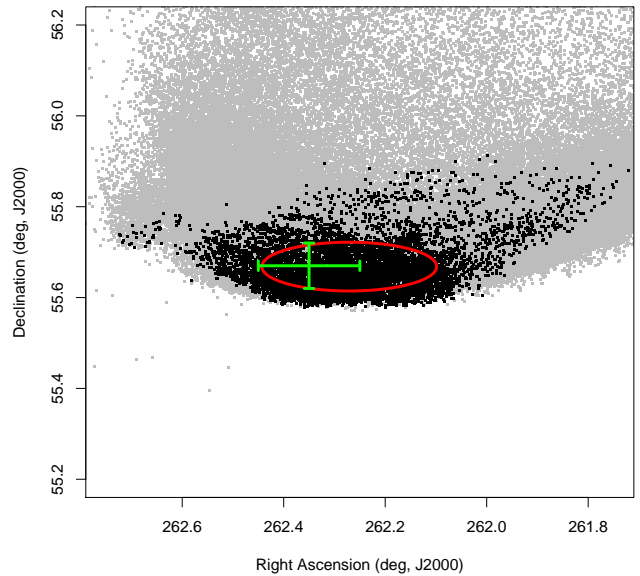}
\caption{The simulated radiant. The grey points indicate those particles passing within $\pm 0.02$~AU and $\pm7$~days of Earth. The black points indicate those meeting the most stringent criteria ($\Delta r = 0.002$~AU, $\Delta t = \pm 1$~day) and are deemed most likely to have comprised the material producing the outburst. The green error bars indicated the CMOR radiant, the red ellipse is the one standard deviation position of the simulated radiant. }
\label{fi:radiant}
\end{figure}

The radiants are shown in Figure~\ref{fi:radiant}. The locations of the CMOR radiant is indicated by the error bars. The ellipse indicates one standard deviation of the simulated radiant. The observed and
simulated radiants thus have $\alpha_g$ and $\delta_g$ which are consistent to within one sigma. The geocentric velocity (far from Earth) of simulated meteors $20.93\pm0.03~\mathrm{km \cdot s^{-1}}$. These agree with the radar-measured velocities within their respective uncertainties.

\subsection{The ``absent'' storm for visual observers}

The size distribution of the outburst in the simulation is concentrated in a narrow range of meteoroid sizes, from about 0.1 to 0.7~mm, corresponding to a typical $\beta \approx 5 \times 10^{-3}$. This size selection arises essentially from the timing constraint. Particles of different sizes tend to have different arrival times at Earth even if released from the parent at the same time and with the same ejection velocity because a particle's orbital period is a function of $\beta$. The need to arrive at the correct time to produce the outburst thus will restrict $\beta$ to a narrow range. This size distribution is qualitatively consistent with CMOR observations. Because of the strong size selection for this outburst, the concept of a power-law distribution of sizes must be used with great care in this case.

The orbital elements for the core of the simulated outburst is listed in Table~\ref{ta:elements}. Given the match in the timing and width of the outburst, as well as consistency between the radar derived sizes and radiants and those of the simulations, we conclude that the 2012 Draconids outburst consisted primarily of particles released during the 1966 perihelion passage of comet 21P/Giacobini-Zinner.

\begin{figure}
\includegraphics[width=0.5\textwidth]{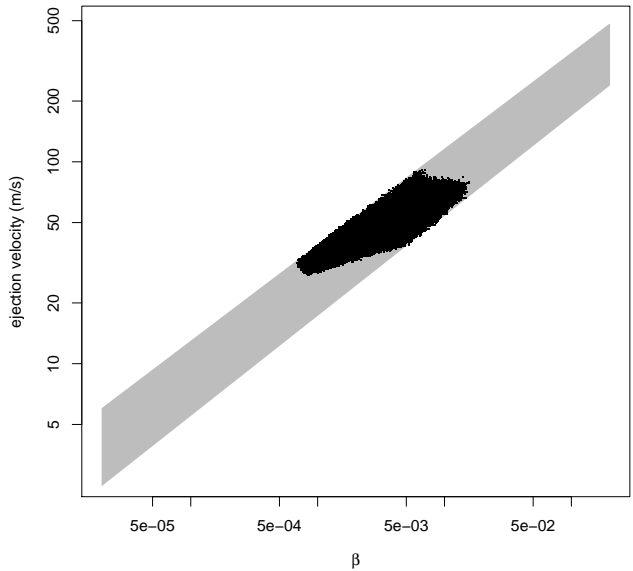}
\caption{The phase space of $\beta$ and ejection speed sampled by the
simulation (grey) and those that are part of the simulated shower (black), demonstrating the reason for dynamical filtering of the outburst to a narrow range of meteoroid sizes. }
\label{fi:betavel}
\end{figure}

The meteoroids which reach the Earth from the 1966 perihelion passage show no concentration of release times in the pre- or post-perihelion orbital arc. The $\beta$ and ejection velocities in the simulation are shown in Figure~\ref{fi:betavel}: only a narrow segment of the particles ejected from the comet reach the Earth during the 2012 outburst.

% \begin{figure}
% \includegraphics[width=\textwidth]{Drac-2012-tvsa.eps}
% \caption{\fix{Quanzhi: you had noted a possible absence of particles with
% certain semimajor axes at certain times (Fig 7 and 8). In the simulations
% there is a variation as well, but it is not quite the same. This plot shows
% the density of a vs solar longitude in the simulations. The black line
% is the radar peak time.}}
% \label{fi:tvsa}
% \end{figure}

The extensive compilation of historical comet observations of \citet{kro99} was examined to see if the 1966 perihelion passage of 21P/Giacobini-Zinner displayed any behaviour that might explain the outburst. The apparition was a faint one and poorly observed but there are no indications of an outburst or other unusual dust production during its 1966 apparition.

\subsection{Future outlook}

We extended our simulations to determine whether any further outbursts of the Draconids are likely between now and 2025. This process is fraught with difficulties, since we have already concluded that our knowledge of the parent's orbit is insufficient to model even the observed 2012 outburst in full detail. However the desire to better understand this peculiar shower leads us to the attempt nonetheless.

Our earlier discussion of the 2012 outburst revealed insufficient accuracy either in our knowledge of the comet's dynamical parameters (i.e. orbital elements, non-gravitational parameters) or in our ability to simulate the comet's evolution. To attempt to get around these restrictions, we run simulations based on each of the sets of osculating orbital elements listed in \citet{marwil08} for all fourteen appearances of 21P/Giacobini-Zinner from its discovery in 1900 through 2005 inclusive. Each is integrated backwards for 100 years, and then forwards again, producing dust at each perihelion passage until the year 2025. The initial simulations and refinement stages are performed in all of these cases as described earlier. In the absence of observation uncertainty and the chaos produced by close encounters with Jupiter, these simulations would all produce the same results. Most -- but not all -- simulations produced similar predictions. However, observational uncertainty and chaos are present and thus each of 
the fourteen orbits is only an approximation to the true orbit at some instant in time, and we find that shower predictions are very sensitive to such small variations in the comet's orbit. The results reported here for future activity represent a somewhat subjective selection of those years where a majority of the simulations produce appreciable activity.

\begin{table*}
 \centering
  \caption{Predicted future Draconid activity.}
  \begin{tabular}{cccccccc}
  \hline
  Year & Trail & Start & Peak  & End    & Max r (mm) & Min r (mm) & Offset?\\
  \hline
  2018 & 1953  & 195.2 & 195.4 & 195.6  & 100        & 0.4        & Yes \\
  2019 & 1979  & 194.0 & 194.2 & 194.5  & 2          & $<0.1$     & No \\
  2021 & 1979  & 193.5 & 193.8 & 194.5  & 0.25       & $<0.1$     & No \\
  2025 & 1933  & 194.0 & 195.0 & 195.3  & 10         & 0.3        & No \\
\hline
\end{tabular}
\label{tbl-future}
\end{table*}

To better understand the situation, we first compared the simulations with known outbursts of the Draconids (2005, 2011, 2012) in the 2001-2012 range. Here we were heartened to able to reproduce the occurrence and absence of outbursts as observed, with one caveat; that the simulated nodal footprint of the 2005 outburst was displaced slightly outside the Earth's orbit similarly to that of 2012 (The 2011 outburst footprint did not require adjustment).

Given these results, we provide here a list of years when the simulations show the possibility of increased Draconids activity where many (though some times not all) the nodal foot prints are on or near the Earth's orbit. These years are 2018, 2019, 2021 and 2025.  The strongest activity is expected from the year 2018 but its nodal footprint is offset slightly from the Earth; a precise activity level is hard to calculate but an outburst similar to 2012 is possible. The other shower years are not offset from the Earth's orbit but show lower rates and smaller particle sizes (Table~\ref{tbl-future}).

\section{Conclusion}

We have reported the analysis of the unexpected intense outburst of the Draconid meteor shower detected by CMOR on October 8, 2012. The peak occurred at $\sim$16:40 UT ($\lambda_{\odot}=195.622^{\circ}$), with a ZHR$_{\mathrm{max}}\approx9000\pm1000$. The weighted mean radiant around the peak time (15--19h UT) was at $\alpha_g=262.4^{\circ}\pm0.1^{\circ}$, $\delta_g=55.7^{\circ}\pm0.1^{\circ}$ (J2000). The mass distribution index was determined to be $s=1.88\pm0.01$, lower than the value found in 2011 indicating that the outburst was dominated by faint meteors compared to 2011. We then used the meteoroid ablation model developed by \citet{cam04} to model the structure of the Draconid meteoroids, which does not reveal unique solutions, but suggests that the number of grains in each meteoroid falls within the range of 10--100.

Visual observations also showed increased activity around the peak time, but with a much lower activity than indicated by radar (ZHR$\sim200$). We intrpret this to indicate a strong size selection in the outburst centred near radar meteoroid sizes, a result consistent with simulations which predict a similar narrow range in small particle sizes. The concept of a single power-law distribution in meteoroid sizes may not be valid for this event spanning radar to visual sizes, indicating that ZHR and flux estimates are similarly uncertain.

Dynamical simulations of this event show with some confidence that this outburst was originated from particles released during the 1966 perihelion passage of the parent body, 21P/Giacobini-Zinner, although there are some uncertainties of the exact timing of the encounter. Further simulations showed that possible increased activity of the Draconid meteor shower may occur in 2018, 2019, 2021 and 2025.

\section*{Acknowledgments}

We thank Dr. David Asher for his very thorough comments and Dr. Rainer Arlt for his helps on the IMO visual data. We also thank Zbigniew Krzeminski, Jason Gill and Daniel Wong for helping with CMOR analysis and operations. Funding support from the NASA Meteoroid Environment Office (cooperative agreement NNX11AB76A) for CMOR operations is gratefully acknowledged. QY thanks 21P/Giacobini-Zinner for giving him a totally unexpected surprise which makes his study and research more cheerful.

Additionally, we thank the following visual observers for their reports which is essential as a confirmation of this outburst: Boris Badmaev, Igor Bukva, Jose Vicente Diaz Martinez, Svetlana Dondokova, Francisco Jose Sevilla Lobato, Ljubica Gra\v{s}i\'c, Ilija Ivanovi\'c, Javor Kac, Julia Mangadaeva, Olga Masheeva, Aleksandar Mati\'c, Jovana Mili\'c, Galina Muhutdinova, Francisco Oca\~na Gonzalez, Rolanda Ondar, Vladimir Osorov, Galina Otto, Sasha Prokofyev, Maciej Reszelski, Alejandro Sanchez De Miguel, Nikolay Solominskiy, Jakub Koukal, Michel Vandeputte, Salvador Aguirre, J\"urgen Rendtel, Terrence Ross, Branislav Savi\'c, Alexandr Maidik, Koen Miskotte, Sne\v{z}ana Todorovi\'c, Sogto Tsurendashiev, Kristina Veljkovi\'c, Allen Zhong, Milena \v{Z}ivoti\'c, and Bimba Zurbaev.

\bibliographystyle{mn2e}
\bibliography{man}

\begin{thebibliography}{}

\bibitem[\protect\citeauthoryear{{Blaauw}, {Campbell-Brown} \&
  {Weryk}}{{Blaauw} et~al.}{2011}]{bla11}
{Blaauw} R.~C.,  {Campbell-Brown} M.~D.,    {Weryk} R.~J.,  2011, \mnras, 412,
  2033

\bibitem[\protect\citeauthoryear{{Borovi{\v c}ka}, {Spurn{\'y}} \&
  {Koten}}{{Borovi{\v c}ka} et~al.}{2007}]{borspukot07}
{Borovi{\v c}ka} J.,  {Spurn{\'y}} P.,    {Koten} P.,  2007, \aap, 473, 661

\bibitem[\protect\citeauthoryear{{Brown} \& {Jones}}{{Brown} \&
  {Jones}}{1995}]{bro95}
{Brown} P.,  {Jones} J.,  1995, Earth Moon and Planets, 68, 223

\bibitem[\protect\citeauthoryear{{Brown}, {Weryk}, {Wong} \& {Jones}}{{Brown}
  et~al.}{2008}]{bro08}
{Brown} P.,  {Weryk} R.~J.,  {Wong} D.~K.,    {Jones} J.,  2008, \icarus, 195,
  317

\bibitem[\protect\citeauthoryear{{Brown} \& {Ye}}{{Brown} \&
  {Ye}}{2012}]{bro12}
{Brown} P.,  {Ye} Q.,  2012, Central Bureau Electronic Telegrams, 3249, 1

\bibitem[\protect\citeauthoryear{{Campbell-Brown}, {Vaubaillon}, {Brown},
  {Weryk} \& {Arlt}}{{Campbell-Brown} et~al.}{2006}]{cam06}
{Campbell-Brown} M.,  {Vaubaillon} J.,  {Brown} P.,  {Weryk} R.~J.,    {Arlt}
  R.,  2006, \aap, 451, 339

\bibitem[\protect\citeauthoryear{{Campbell-Brown} \&
  {Koschny}}{{Campbell-Brown} \& {Koschny}}{2004}]{cam04}
{Campbell-Brown} M.~D.,  {Koschny} D.,  2004, \aap, 418, 751

\bibitem[\protect\citeauthoryear{{Ceplecha}, {Borovi{\v c}ka}, {Elford},
  {Revelle}, {Hawkes}, {Porub{\v c}an} \& {{\v S}imek}}{{Ceplecha}
  et~al.}{1998}]{cep98}
{Ceplecha} Z.,  {Borovi{\v c}ka} J.,  {Elford} W.~G.,  {Revelle} D.~O.,
  {Hawkes} R.~L.,  {Porub{\v c}an} V.,    {{\v S}imek} M.,  1998, \ssr, 84, 327

\bibitem[\protect\citeauthoryear{{Crifo} \& {Rodionov}}{{Crifo} \&
  {Rodionov}}{1997}]{crirod97}
{Crifo} J.~F.,  {Rodionov} A.~V.,  1997, Icarus, 127, 319

\bibitem[\protect\citeauthoryear{{Davidson}}{{Davidson}}{1915}]{dav15}
{Davidson} R.~M.,  1915, JBAA, 25, 292

\bibitem[\protect\citeauthoryear{{Denning}}{{Denning}}{1926}]{den27}
{Denning} W.~F.,  1926, \mnras, 87, 104

\bibitem[\protect\citeauthoryear{{Everhart}}{{Everhart}}{1985}]{eve85}
{Everhart} E.,  1985, in {Carusi} A.,  {Valsecchi} G.~B.,  eds, Dynamics of
  Comets: Their Origin and Evolution {An efficient integrator that uses
  Gauss-Radau spacings}.
Kluwer, Dordrecht, pp 185--202

\bibitem[\protect\citeauthoryear{{Hocking}, {Fuller} \& {Vandepeer}}{{Hocking}
  et~al.}{2001}]{hoc01}
{Hocking} W.~K.,  {Fuller} B.,    {Vandepeer} B.,  2001, Journal of Atmospheric
  and Solar-Terrestrial Physics, 63, 155

\bibitem[\protect\citeauthoryear{{Jacchia}, {Kopal} \& {Millman}}{{Jacchia}
  et~al.}{1950}]{jac50}
{Jacchia} L.~G.,  {Kopal} Z.,    {Millman} P.~M.,  1950, \apj, 111, 104

\bibitem[\protect\citeauthoryear{{Jenniskens}}{{Jenniskens}}{1995}]{jen95}
{Jenniskens} P.,  1995, \aap, 295, 206

\bibitem[\protect\citeauthoryear{{Jenniskens}}{{Jenniskens}}{2006}]{jen06}
{Jenniskens} P.,  2006, {Meteor Showers and their Parent Comets}

\bibitem[\protect\citeauthoryear{{Jenniskens}, {Barentsen} \&
  {Yrjola}}{{Jenniskens} et~al.}{2011}]{jen11}
{Jenniskens} P.,  {Barentsen} G.,    {Yrjola} I.,  2011, Central Bureau
  Electronic Telegrams, 2862, 1

\bibitem[\protect\citeauthoryear{{Jones}}{{Jones}}{1995}]{jon95}
{Jones} J.,  1995, \mnras, 275, 773

\bibitem[\protect\citeauthoryear{{Jones}, {Brown}, {Ellis}, {Webster},
  {Campbell-Brown}, {Krzemenski} \& {Weryk}}{{Jones} et~al.}{2005}]{jon05}
{Jones} J.,  {Brown} P.,  {Ellis} K.~J.,  {Webster} A.~R.,  {Campbell-Brown}
  M.,  {Krzemenski} Z.,    {Weryk} R.~J.,  2005, \planss, 53, 413

\bibitem[\protect\citeauthoryear{{Kero}, {Fujiwara}, {Abo}, {Szasz} \&
  {Nakamura}}{{Kero} et~al.}{2012}]{ker12}
{Kero} J.,  {Fujiwara} Y.,  {Abo} M.,  {Szasz} C.,    {Nakamura} T.,  2012,
  \mnras, 424, 1799

\bibitem[\protect\citeauthoryear{{Koschack} \& {Rendtel}}{{Koschack} \&
  {Rendtel}}{1990}]{kos90}
{Koschack} R.,  {Rendtel} J.,  1990, WGN, Journal of the International Meteor
  Organization, 18, 44

\bibitem[\protect\citeauthoryear{{Koten}, {Vaubaillon}, {Toth}, {Zenden},
  {McAuliffe}, {Koschny} \& {Pautet}}{{Koten} et~al.}{2012}]{kot12}
{Koten} P.,  {Vaubaillon} J.,  {Toth} J.,  {Zenden} J.,  {McAuliffe} J.,
  {Koschny} D.,    {Pautet} D.,  2012, LPI Contributions, 1667, 6225

\bibitem[\protect\citeauthoryear{Kronk}{Kronk}{1999}]{kro99}
Kronk G.~W.,  1999, Cometography.
Vol.~1, Cambridge University Press, Cambridge

\bibitem[\protect\citeauthoryear{Marsden, Sekanina \& Yeomans}{Marsden
  et~al.}{1973}]{marsekyeo73}
Marsden B.~G.,  Sekanina Z.,    Yeomans D.~K.,  1973, 78, 211

\bibitem[\protect\citeauthoryear{Marsden \& Williams}{Marsden \&
  Williams}{2008}]{marwil08}
Marsden B.~G.,  Williams G.~V.,  2008, Catalogue of Cometary Orbits, 17$^{th}$
  edn.
{IAU} Central Bureau for Astronomical Telegrams -- Minor Planet Center,
  Cambridge, Massachusetts

\bibitem[\protect\citeauthoryear{{Maslov}}{{Maslov}}{2011}]{mas11}
{Maslov} M.,  2011, WGN, Journal of the International Meteor Organization, 39,
  64

\bibitem[\protect\citeauthoryear{{McIntosh}}{{McIntosh}}{1968}]{mci68}
{McIntosh} B.~A.,  1968, in {Kresak} L.,  {Millman} P.~M.,  eds, Physics and
  Dynamics of Meteors Vol.~33 of IAU Symposium, {Meteor Mass Distribution from
  Radar Observations}.
p.~343

\bibitem[\protect\citeauthoryear{{McKinley}}{{McKinley}}{1961}]{mck61}
{McKinley} D.~W.~R.,  1961, {Meteor science and engineering.}

\bibitem[\protect\citeauthoryear{{Mikkola} \& {Innanen}}{{Mikkola} \&
  {Innanen}}{1999}]{mikinn99}
{Mikkola} S.,  {Innanen} K.,  1999, 74, 59

\bibitem[\protect\citeauthoryear{{Sato}, {Watanabe} \& {Ohkawa}}{{Sato}
  et~al.}{2012}]{sat12}
{Sato} M.,  {Watanabe} J.,    {Ohkawa} T.,  2012, LPI Contributions, 1667, 6319

\bibitem[\protect\citeauthoryear{Standish}{Standish}{1998}]{sta98}
Standish E.~M.,  1998, Technical report, Planetary and Lunar Ephemerides
  {DE405/LE405}.
NASA Jet Propulsion Laboratory

\bibitem[\protect\citeauthoryear{{Tancredi}, {Fern{\'a}ndez}, {Rickman} \&
  {Licandro}}{{Tancredi} et~al.}{2000}]{tan00}
{Tancredi} G.,  {Fern{\'a}ndez} J.~A.,  {Rickman} H.,    {Licandro} J.,  2000,
  \aaps, 146, 73

\bibitem[\protect\citeauthoryear{{Vaubaillon}, {Koten}, {Bouley}, {Rudawska},
  {Maquet}, {Colas}, {Toth}, {Zender}, {McAuliffe}, {Pautet}, {Koschny},
  {Jenniskens}, {Leroy}, {Lecacheux} \& {AntierHe}}{{Vaubaillon}
  et~al.}{2012}]{vau12}
{Vaubaillon} J.,  {Koten} P.,  {Bouley} S.,  {Rudawska} R.,  {Maquet} L.,
  {Colas} F.,  {Toth} J.,  {Zender} J.,  {McAuliffe} J.,  {Pautet} D.,
  {Koschny} D.,  {Jenniskens} P.,  {Leroy} A.,  {Lecacheux} J.,    {AntierHe}
  K.,  2012, LPI Contributions, 1667, 6280

\bibitem[\protect\citeauthoryear{{Vaubaillon}, {Watanabe}, {Sato}, {Horii} \&
  {Koten}}{{Vaubaillon} et~al.}{2011}]{vau11}
{Vaubaillon} J.,  {Watanabe} J.,  {Sato} M.,  {Horii} S.,    {Koten} P.,  2011,
  WGN, Journal of the International Meteor Organization, 39, 59

\bibitem[\protect\citeauthoryear{{Weidenschilling} \&
  {Jackson}}{{Weidenschilling} \& {Jackson}}{1993}]{weijac93}
{Weidenschilling} S.~J.,  {Jackson} A.~A.,  1993, Icarus, 104, 244

\bibitem[\protect\citeauthoryear{{Weryk} \& {Brown}}{{Weryk} \&
  {Brown}}{2012}]{wer12}
{Weryk} R.~J.,  {Brown} P.~G.,  2012, \planss, 62, 132

\bibitem[\protect\citeauthoryear{{Wiegert}, {Vaubaillon} \&
  {Campbell-Brown}}{{Wiegert} et~al.}{2009}]{wievaucam09}
{Wiegert} P.,  {Vaubaillon} J.,    {Campbell-Brown} M.,  2009, Icarus, 201, 295

\bibitem[\protect\citeauthoryear{{Wiegert}, {Brown}, {Weryk} \&
  {Wong}}{{Wiegert} et~al.}{2013}]{wiebrower13}
{Wiegert} P.~A.,  {Brown} P.~G.,  {Weryk} R.~J.,    {Wong} D.~K.,  2013, \aj,
  145, 70

\bibitem[\protect\citeauthoryear{{Ye}, {Brown}, {Campbell-Brown} \&
  {Weryk}}{{Ye} et~al.}{2013}]{ye13}
{Ye} Q.,  {Brown} P.~G.,  {Campbell-Brown} M.~D.,    {Weryk} R.~J.,  2013,
  \mnras, 436, 675

\bibitem[\protect\citeauthoryear{Yeomans}{Yeomans}{1986}]{yeo86}
Yeomans D.~K.,  1986, Quart. J. Roy. Astron. Soc., 27, 116

\end{thebibliography}

\label{lastpage}

\end{document}